\documentstyle[12pt]{article}

\makeatletter
\@addtoreset{equation}{section}
\makeatother

\begin{document}
\hfill{}
\hfill{}

\hfill{UCSBTH-97-07}

\hfill{hep-th/9704204}

\vspace{24pt}

\begin{center}
{\large {\bf Black Holes as Fundamental Strings:
Comparing the Absorption of Scalars}}

\vspace{48pt}

Roberto Emparan

\vspace{12pt}

{\sl Department of Physics}\\
{\sl University of California}\\
{\sl Santa Barbara, CA 93106}\\
{\it emparan@cosmic1.physics.ucsb.edu}

\vspace{72pt}

{\bf Abstract}
\end{center}
\begin{quote}{\small
The recently proposed ``correspondence principle'' of Horowitz and
Polchinski provides a concrete means to relate (among others)
black holes with
electric NS-NS charges to fundamental strings and correctly match
their entropies. We test further this correspondence by examining
the greybody factors in the absorption rates of neutral,
minimally coupled scalars by a near extremal black hole.
Perhaps surprisingly, the results disagree in general
with the absorption by weakly coupled strings. Though this does not
disprove the correspondence, it indicates that it might not be
simple in this region of the black hole parameter space.}
\end{quote}

\newpage

\section{Introduction}

During the past year there has been impressive progress in our
understanding of the microscopic description of black holes
\cite{strominger} (see
\cite{horowitz,maldacena} for reviews and
further references). For the case that is best understood, 
a description has emerged
for the weak coupling dynamics
of a five dimensional black hole
near extremality in terms of an ``effective string''.
The latter is in fact a bound state of D-strings
restricted to move inside
a D-fivebrane, and
which is excited above the BPS state by having both left and
right moving momenta running along the string (in a
dilute gas regime) \cite{callan,hs}. This model
precisely reproduces the Bekenstein-Hawking
entropy of the black hole in terms of the degeneracy of states
of the effective string, and
Hawking radiation comes about as the emission of
closed string states resulting from annihilation of pairs of
left and right moving quanta.
This effective description has
turned out to be surprisingly successful.
Not only the entropy can be correctly
reproduced, but also
details of the scalar emission rates
computed from the effective string
show precise agreement with the spectrum of Hawking radiation.
This agreement is correct including normalization factors \cite{dmtwo}
and extends to the level of black hole greybody profiles
\cite{ms}.
\footnote{The agreement has been further reinforced in
\cite{klebetal}.
However, it should be noted that disagreement has also been found
in certain specific regimes \cite{fay,taylor,dealwis}.}
A possible explanation for this success has been proposed in
\cite{maldalow}.

Still within the near-extremal realm, it has been
shown in \cite{maldafive} that when one adds D-string-antistring
pairs to the bound state described above,
the black hole entropy can be correctly accounted for by
the states of a non-critical Polyakov string with its
effective tension
and central charge constrained by
the D-fivebrane background. In
this regime, however, it is
unclear how strong coupling effects can be avoided.
Moreover, the details of spectroscopy show only partial agreement,
and deviations occur at the level
of leading order corrections to the absorption cross section at
low energies \cite{km}.

Pushing the picture further along these lines, one could expect
to find a description of the (five dimensional) neutral
Schwarzschild black hole
as a bound state of a number of
D-fivebranes, D-strings and momenta together with an equal number
of their anti-excitations \cite{hms}.
However, it is not clear at all how to reliably compute the
degeneracy of states for such a system,
or even to justify or understand, in general,
why the brane-antibrane pairs should not annihilate.

A rather different approach for understanding
the entropy of the Schwarzschild black hole was
initiated by Susskind a few years back \cite{susskind}. Here,
the working hypothesis is that
the only objects needed to account for the degrees of freedom of a
neutral black hole are fundamental strings ---and not, in particular,
D-branes. In this picture, if we start at
strong coupling with a state that looks like a black hole, then
as we decrease the coupling the horizon shrinks.
Eventually, the state is better described as a long,
highly excited string. This conjecture has been carried further in
\cite{halyo1,stanford,hp}.\footnote{See \cite{larsen} for another,
black-hole-guided, 
approach to a string model for black holes arbitrarily
far from extremality.} The obvious
difficulty with this approach, namely, the different growth of
the number of states with a given energy for black holes and strings,
has been recently solved in a convincing manner by Horowitz and
Polchinski in \cite{hp}.
The key observation is that we should not expect
the mass of a certain black hole state ---which is fixed
in Planck units--- to be equal to that
of the corresponding string state ---constant in string units---
for arbitrary values of the string coupling constant $g$.
Rather, the
correspondence between black hole and string parameters
should be naturally made at the value of $g$ for which
the transition between both descriptions takes place.
There is a shortcoming here in that we do not know how to determine
precisely at which value of the coupling
this transition happens. Hence this principle only allows to relate
quantities up to factors of order one. Nevertheless, its
range of applicability is wider than for other approaches
---most of which it
subsumes---, and it has been
shown in \cite{hp} to yield the correct dependence of the
entropy on black hole parameters for a
large number of non trivial cases, such as, in particular,
black holes with charges corresponding to the winding and momentum
numbers of a fundamental string.

A key premise of the Horowitz-Polchinski
correspondence principle
is that, when the strongly coupled black hole and the weakly
coupled string are taken to the transition point, the mass of
neither of them
changes by a large factor. In practice, this means that, at the
matching point, the classical black hole mass can be taken to be
approximately equal to the energy of a string whose levels are
determined
according to the spectrum of a {\it free} (or very weakly coupled)
string. Neither the finite coupling corrections to the string mass
spectrum, nor the string-size modifications to the black hole
geometry introduce large factors that could alter the matching
of masses and entropies at the transition point. Our aim in this paper
is to test if this simplest
classical-black-hole/free-string correspondence is enough to
account for details of the radiation such as greybody filtering.
This is particularly
interesting because, as was revealed in \cite{ms}, greybody
factors encode information on the excitation 
spectrum of the string. For technical reasons, 
the computations in the black hole side will
have to be restricted to near extremal configurations in five and
four dimensions. The results
we find indicate that, at least in the region of the
black hole parameter
space that we are able to probe, the simplest model for the
correspondence does not seem to account correctly for the
details of the spectrum. There is some evidence that
the discrepancy could be traced to the fact that, for the class of
black holes under consideration, the near 
extremal state at the matching point 
actually is {\it not} infinitesimally close to 
the BPS limit.

The paper is organized as follows. In Sec.\ \ref{s5bh} we review the
calculation of scalar absorption by 5D and 4D
near extremal black holes which
will be of use later on. Then, in Sec.\ \ref{fsvsbh} we make use of the
correspondence principle to compare these results
with the absorption rates of neutral scalars by fundamental strings
previously computed in \cite{stanford}. We give a simple argument
which shows that, quite generally,
disagreement is
found. We conclude in Sec.\ \ref{concl} with a discussion of the
results and their implications.

\section{Absorption rates for near extremal five and four
dimensional black holes}\label{s5bh}

The absorption cross sections, including greybody factors,
of minimally coupled scalars by
a near extremal black hole in five and four dimensions
have been computed
in great generality in \cite{km}. 
These results will be needed later, so we find 
it useful to describe how they are obtained. 

We start with the non-rotating non-extremal five dimensional
black hole solution of string theory \cite{cy,hms}
which takes the form, in Einstein
conformal gauge,
\begin{equation}\label{5dbh}
ds^2 = -{h\over f^{2/3}} dt^2 + f^{1/3} ({dr^2\over h} +
r^2 d\Omega_3^2),
\end{equation}
with
\begin{eqnarray}\label{pots}
h &=& 1-{r_0^2\over r^2},\qquad f=f_1 f_2 f_3 ,\nonumber\\
f_i &=& 1+{r_i^2\over r^2},\quad r_i^2 = r_0^2 \sinh^2\sigma_i ,
\quad i= 1,2,3.
\end{eqnarray}
This solution admits a variety of embeddings in several different
compactifications of any of the superstring (and M-) theories.
The characteristic radii $r_i$
are associated with three different types of charges, and their
interpretation (as, e.g., KK momentum, winding, RR electric or
magnetic charge,
etc), depends on the embedding. In particular,
for the main purposes of this
paper (Sec.\ \ref{fsvsbh}) we shall only need to consider two
non-zero charges.
Nonetheless, for the moment we
leave the number of charges and their interpretation otherwise
unspecified,
but take the radii
to be ordered as
\begin{equation}
r_1 \geq r_2 \geq r_3.
\end{equation}
Besides the ``non-extremality
parameter'' $r_0$,
one works
with the three radii,
$r_i$, or alternatively, the associated hyperbolic angles,
$\sigma_i$, depending
on convenience.

The mass, entropy and Hawking temperature of the black hole are
\begin{eqnarray}\label{5dmass}
M &=& {\pi r_0^2\over 8 G_5} (\cosh 2\sigma_1+ \cosh 2\sigma_2+
\cosh 2\sigma_3),\nonumber\\
S &=& {\pi^2 r_0^3\over 2 G_5}\cosh\sigma_1\cosh\sigma_2\cosh\sigma_3,
\\
T_H^{-1} &=& 2\pi r_0\cosh\sigma_1\cosh\sigma_2\cosh\sigma_3.
\nonumber
\end{eqnarray}
The black hole is taken to near extremality
by having {\it at least}
one large charge, say $r_0 \ll  r_1$, so that
$r_1 = r_0 \sinh\sigma_1
\approx r_0 \cosh\sigma_1$. The other two radii (charges)
can be as well large or small.
The scattering will be restricted to
s-waves, higher partial waves being
suppressed
by centrifugal potential barriers at
frequencies such that $\omega r_0\ll 1$ and $\omega r_1 < 2$
\cite{km}.
Although
it would not be difficult
to include higher angular momenta,
our primary interest will be on a range of low enough
frequencies specified by $\omega r_1\ll 1$.

The wave equation for a minimally coupled,
spherically symmetric scalar field
$\Phi(r,t) = e^{-i\omega t} R(r)$ in this black hole
background takes
the form
\begin{equation}\label{weq1}
\left({h\over r^3}{d\over dr} h r^3{d\over dr} +
\omega^2 f\right)R=0.
\end{equation}

An alternative set of four
parameters characterizing the black hole is
useful when dealing with the wave equation. These are
\begin{eqnarray}\label{dcef}
D &=& {\omega^2\over 4 r_0^4} r_1^2 r_2^2 r_3^2,\nonumber\\
C &=& {\omega^2\over 4 r_0^2} (r_1^2 r_2^2 + r_1^2 r_3^2 +
 r_2^2 r_3^2),
\nonumber\\
E &=& {\omega^2\over 4} (r_1^2 + r_2^2 + r_3^2),\\
E_2 &=& {\omega^2\over 4} r_0^2.\nonumber
\end{eqnarray}

It is also convenient to define the variable $z=h(r)$
in terms of which the wave equation takes the form
\begin{equation}\label{weqdcef}
\left(z{d\over dz}\right)^2 R + \bigg( D + {C\over (1-z)}+
{E\over (1-z)^2}+
{E_2\over (1-z)^3}\bigg)  R =0.
\end{equation}
The latter is still the exact wave equation for the s-wave scalar.
Since, in general, it is not possible to solve it analytically
for arbitrary values of $r$,
one needs to resort to approximate methods.\footnote{However,
it must be noted that for the extremal
black hole with {\it only} one single type of charge it is indeed
possible to analytically solve
the wave equation exactly, everywhere and for arbitrary frequencies
(and also include higher partial waves),
in terms of Bessel functions. For this black hole, 
though, the horizon
is a singularity of zero area. In a similar way, for the
extreme four dimensional black hole with two charges the exact
solution can be
found in terms of Coulomb wave functions.}
The traditional way to deal with the problem
\cite{star,press,page,unruh,gibbons,dmtwo,ms}
has been to, first, solve the
equations in two different regions:
(I) near the horizon; (II) far from the black hole, 
and then match the solutions at some point
inbetween, or by means of an intermediate region.
For near extremal black holes of the sort under
consideration, Rajaraman has studied in detail
the definition of regions (I) and (II),
and has settled the question of the
existence of an overlap region and the matching 
of solutions \cite{rajaraman}.

Boundary conditions are imposed at the black hole horizon,
by requiring the wave to be purely ingoing into the black hole.
The matching
of solutions in (I) and (II) then allows to determine
the ratio at
asymptotic infinity of
the amplitudes of radially ingoing waves to those outcoming,
and hence the luminosity of the black hole.

Very close to the horizon $(z\rightarrow 0)$
an approximate solution can be found if we neglect the $z$ dependence
in the non-derivative term in (\ref{weqdcef}).
In this way one finds
\begin{equation}\label{RIo}
R_I \simeq e^{\pm i \sqrt{D+C+E+E_2}\ln z}=\exp\left(\pm {i\omega
A_{bh}\over 4\pi^2 r_0^2}\ln z\right),
\end{equation}
where the $+$ $(-)$ solution is outcoming (ingoing) at the horizon.
This is essentially the form of the solution used in \cite{unruh,page},
and
can be seen to correctly yield the leading term, at low frequencies,
of the
absorption cross section as equal to
the horizon area of the black hole. However,
in order to keep further dependence on $z$
and be
able to enlarge the distance within which the solution is valid,
it is found adequate \cite{ms,km} to try the ansatz
$R_I=z^\alpha (1-z)^\beta F(z)$.
In the region where $E_2$ can be dropped
a solution can be found with 
$F$ a hypergeometric function.
Explicitly,
\begin{equation}\label{RI}
R_I = z^\alpha (1-z)^\beta F(\alpha+\beta +i\sqrt{D},
\alpha+\beta -i\sqrt{D};1+2\alpha;z)
\end{equation}
with
\begin{equation}\label{alpbet2}
\alpha = -i\sqrt{D+C+E},\qquad
\beta= {1\over 2}(1-\sqrt{1-4E}).
\end{equation}
Neglection of the $E_2$ term is certainly 
justified provided $r\ll r_1$.

The solution has been completely fixed, 
up to arbitrary global wave function normalization,
by demanding that, very close to the horizon,
the solution behaves as the ingoing wave (\ref{RIo}).
We also need the limiting form of $R_I$ 
for large $r$, i.e.,
$z\rightarrow 1$. This is
\begin{equation}\label{nearlarge}
R_I\sim (1-z)^\beta{\Gamma(1+2\alpha)\;\Gamma(1-2\beta)\over
\Gamma(1+ \alpha-i\sqrt{D} -\beta)\;
\Gamma(1+ \alpha+i\sqrt{D} -\beta)}.
\end{equation}

To analyze the equation far from the horizon it is
convenient to define $\rho=\omega r$ and 
$R=\psi(\rho)/\rho$. If $r\gg r_0$ we can approximate
$h\simeq 1$. The wave equation becomes
\begin{equation}
\rho^2\psi'' +\rho\psi' - (1-\rho^2 f)\psi=0.
\end{equation}
Approximate now $1-\rho^2 f\approx 1-4E -\rho^2$, 
which requires $1 \gg \omega^4 r_1^2 r_2^2/\rho^2$, 
i.e., 
$r\gg \omega r_1 r_2$. 
Upon doing so we find a
Bessel equation for $\psi$, 
and the general solution can 
be expressed in terms of Bessel functions
$J_{\nu}(\omega r)$ as
\begin{equation}\label{RII}
R_{II} = {A\over \omega r} J_{1-2\beta}(\omega r)
+{B\over \omega r} J_{2\beta-1}(\omega r).
\end{equation}

{}For small $\rho$
the behavior of $R_{II}$ is
found from
\begin{equation}\label{farshort}
{1\over \rho} J_{1-2\beta}(\rho)\sim {1\over 2}
\left({2\over \rho}\right)^{2\beta} {1\over \Gamma(2-2\beta)},
\end{equation}
\begin{equation}\label{farshort2}
{1\over \rho} J_{2\beta-1}(\rho)\sim {2\over \rho^2}
\left({\rho\over 2}\right)^{2\beta} {1\over \Gamma(2\beta)}.
\end{equation}
{}From here we see that
the wave $J_{1-2\beta}(\omega r)/(\omega r)$ behaves,
in this region in exactly
the same way as the wave $R_I$ for $r\gg r_0$, 
(\ref{nearlarge}).
Moreover, $R_I$ is definitely valid up to $r\ll r_1$
(in fact, up to $r\ll 1/\omega$ \cite{rajaraman}),
and $R_{II}$ down to $r\gg\omega r_1 r_2$. 
Hence, there exists an overlap region
where the matching can
be done. The coefficients in (\ref{RII}) are determined
as $B=0$ \footnote{More precisely, 
$B/A$ is small when $\beta$ is small. 
We thank J.~Traschen for discussions on this point.} and
\begin{equation}\label{matcha}
A = 2\left({\omega r_0\over 2}\right)^{2\beta}
{\Gamma(1+2\alpha)\;\Gamma(1-2\beta)\;\Gamma(2-2\beta)\over
\Gamma(1+ \alpha-i\sqrt{D} -\beta)\;\Gamma(1+ \alpha+i\sqrt{D} 
-\beta)}.
\end{equation}
This fixes, up to global wave function
normalization, the solution to the wave equation with the required
boundary conditions at the horizon.

Having constructed this solution for s-wave scattering, and in the
approximation
where we neglect higher partial waves,
the plane wave absorption cross section
can be finally found to be
\begin{equation}
\sigma_{\rm abs} = {16\pi^2 i\alpha r_0^2\over \omega |A|^2}.
\end{equation}
This form of the solution was first found in \cite{km}.

The expression for $\sigma_{\rm abs}$ is 
much more amenable to
physical interpretation if $|A|^2$ is expanded
for small $\beta$.
Still, since
\begin{equation}
i\alpha \sim \max\left\{ \omega r_1,
{\omega r_1 r_2\over r_0},{\omega r_1 r_2 r_3\over r_0^2}\right\},
\end{equation}
it is possible to keep $|\alpha|$ of order one as long as
there are, at least, two large charges \footnote{But keep in mind
that they need not be of the same order, i.e., we can
have $r_1\gg r_2 \gg r_0$.}.
In this regime (which is the one analyzed in 
\cite{ms}) straightforward algebra yields
\begin{equation}\label{sigma1}
 \sigma_{\rm abs} = {4\pi^3 r_0^2 \;|\alpha^2 +D|\over \omega}
{e^{\;4\pi i\alpha} -1\over  (e^{\;2\pi i(\alpha+
i\sqrt{D})}-1)
(e^{\;2\pi i(\alpha-i\sqrt{D})}-1)}
[1 + O(\omega^2 r_1^2)].
\end{equation}

If there is just one large charge,
then not only $\beta$, but also $|\alpha|$ and $\sqrt{D}$ have 
to be regarded as
very small.
The absorption cross section should be written now as
\begin{equation}\label{sigma2}
 \sigma_{\rm abs} = {4\pi^2 r_0^2\; i\alpha\over \omega}
\bigg[1 + {\pi^2\over 3}  |\alpha^2 +D|
- 4 \beta\left(\ln({\omega r_0\over2}) -1 +\gamma\right) +
O(\omega^4 r_1^4) \bigg],
\end{equation}
where $\gamma$ is the Euler-Mascheroni constant.
The logarithmic term comes from
expanding the term $(\omega r_0/2)^{4\beta}$ in $|A|^2$.

Let us rewrite these expressions
for $\sigma_{\rm abs}$ in terms of physical quantities.
Given that we always assume that
there is at least one large charge,
one has
\begin{eqnarray}\label{ialpha}
i\alpha &=& {\omega A_{bh}\over 4\pi^2 r_0^2}
[1+O(r_0^2/r_1^2)]\nonumber\\
&=& {\omega\over 4\pi T_H}[1+O(r_0^2/r_1^2)].
\end{eqnarray}
{}From here and (\ref{sigma2}) one finds that,
for $\omega\rightarrow 0$, the absorption cross section is strictly
equal to the black hole area,
$\sigma_{\rm abs}= A_{bh} + O(\omega)$ \cite{unruh,dgm}.

Consider now small but non-vanishing frequencies. From
the second expression for $i\alpha$ in (\ref{ialpha}) we see that
the factor
$(e^{\;4\pi i\alpha} -1)$ in (\ref{sigma1}) is precisely
the Planckian factor for the Hawking radiation.
This suggests to define,
in a similar way,
two other ``left'' and ``right''
temperatures as
\begin{eqnarray}\label{lrtemps}
T_{R,L}^{-1} &=& {4\pi\over \omega}(i\alpha \pm \sqrt{D})\nonumber\\
&\simeq& 2\pi r_1\cosh(\sigma_2\pm\sigma_3 ).
\end{eqnarray}
The last expression is found bearing in mind that $r_1$ is a large
radius.
The absorption cross section (\ref{sigma1}) presents now the
suggestive form
\begin{equation}\label{sigmaphys}
 \sigma_{\rm abs} = A_{bh}\;{\omega\over 2(T_R + T_L)}
{e^{\omega\over T_H} -1\over  (e^{\omega\over 2 T_R}-1)
(e^{\omega\over 2 T_L}-1)}
[1 + O(\omega^2 r_1^2)].
\end{equation}
On the other hand, if there is only one large charge,
$(\omega/T_{L,R,H})^2\sim E\ll 1$,
the result is
\begin{equation}\label{sigmaphys2}
 \sigma_{\rm abs} = A_{bh}
\bigg[1 + {\omega^2\over 48}{1\over T_R T_L}
- \omega^2 r_1^2\left(\ln({\omega r_0\over2}) 
-1 +\gamma\right) +
O(\omega^4 r_1^4)\bigg].
\end{equation}
Notice that the most
relevant correction term is logarithmic,
$\sim \beta\ln(\omega r_0)$.
Since $\omega r_0\ll 1$, this can be quite significant and
is larger than the terms coming from the exponential
greybody factors.

The greybody factors of thermal form in (\ref{sigmaphys})
were found by Maldacena and Strominger \cite{ms},
who also showed that
$T_{R,L}$ correspond exactly to effective temperatures for the
left and right moving vibrations of the
``effective string'' model for the five-dimensional black hole
with two large D-brane charges.
The case when just one of the charges
is large has been solved more recently in \cite{km}, and the
logarithmic correction noted. Its absence from the
simplest string calculation has been interpreted as suggesting
that the effective
string model should be modified
in the corresponding region of parameter space.

In four dimensions we consider the
non-rotating non-extremal
black hole with four charges,
\begin{equation}\label{4dbh}
ds^2 = -{h\over f^{1/2}} dt^2 + f^{1/2} ({dr^2\over h} +
r^2 d\Omega_2^2),
\end{equation}
with
\begin{eqnarray}\label{4pots}
h &=& 1-{r_0\over r},\qquad f=f_1 f_2 f_3 f_4,\nonumber\\
f_i &=& 1+{r_i\over r},\quad r_i = r_0 \sinh^2\sigma_i ,
\quad i= 1,2,3,4,
\end{eqnarray}
and mass, entropy and temperature,
\begin{eqnarray}
M &=& {r_0\over 8 G_4} \sum_{i=1}^4\cosh 2\sigma_i,\nonumber\\
S &=& {\pi r_0^2\over G_4}\prod_{i=1}^4\cosh \sigma_i,
\\
T_H^{-1} &=& 4\pi r_0\prod_{i=1}^4\cosh \sigma_i.
\nonumber
\end{eqnarray}
As before, we order the radial parameters as
\begin{equation}
r_1 \geq r_2 \geq r_3 \geq r_4.
\end{equation}

The analysis of absorption rates, carried out in \cite{km},
is quite similar to that for the 5D black hole.
However, the 4D case differs from the 5D case in two crucial
respects: first, in
order to be able to solve the equations one
needs to restrict to the case with {\it at least two}
large charges,
\begin{equation}
r_1,r_2 \gg r_0.
\end{equation}
Thus, we are not able to deal with the most general near extremal
black holes in four dimensions.

Second, the leading correction at low frequencies is found to be
linear in $\omega$, in contrast to the 5D result where
corrections start to appear at order $\omega^2$.
As a consequence, at low enough frequencies the linear term
completely
masks the (quadratic) terms coming from expanding the exponential
thermal factors, as well as
the logarithmic
correction terms found in the previous section.
More importantly, such
correction (also found, in a slightly different context, in
\cite{dasgupta}) does not seem to appear in the results
obtained using the
effective string model for the 4D black hole \cite{km}.

Near the horizon
the analysis is very closely similar to that in 
five dimensions.
On the other hand, far from the horizon the wave 
equation
can be solved in terms of Coulomb wave functions,
$F_L(\eta,\omega r)$, (with non-integer $L$) as
\begin{equation}\label{4RII}
R_{II} = {A\over \omega r} F_{-\beta}(\eta,\omega r)
+{B\over \omega r} F_{\beta-1}(\eta,\omega r),
\end{equation}
where
\begin{equation}
\eta \equiv -{\omega \over 2}
(r_1+r_2+r_3+r_4),
\end{equation}
and $\beta\approx \omega^2 r_1 r_2$ are small quantities.

{}For
small $\omega r$ the functions $F_L(\eta,\omega r)$
have again a power-like dependence
that enables $R_{II}$ in (\ref{4RII})
to be easily matched to the near
region solution (\ref{nearlarge}).
Define now
\begin{eqnarray}
T_{R,L}^{-1} &=&  4\pi \sqrt{r_1 r_2}\cosh(\sigma_3\pm\sigma_4 ).
\end{eqnarray}
When $\omega/T_{L,R,H}\sim 1$ (which now requires at
least three large charges),
the absorption cross section
takes the form
\begin{equation}\label{4sigmaphys}
\sigma_{\rm abs} = A_{bh}\;{\omega\over 2(T_R + T_L)}
{e^{\omega\over T_H} -1\over  (e^{\omega\over 2 T_R}-1)
(e^{\omega\over 2 T_L}-1)}
[1 + O(\omega r_1)].
\end{equation}

When there are only two large charges
$\omega/T_{R,L}$ are small, 
and the result would be
\begin{equation}\label{4sigmaphys2}
 \sigma_{\rm abs} = A_{bh}
\left(1 - \pi\eta + {\omega^2\over 48}{1\over T_R T_L} +\dots
 \right).
\end{equation}
Since $\eta$ is already linear in $\omega r_1$ we have neglected
the terms proportional to $\beta$, similar to those found in five
dimensions.
Indeed, given that $\omega^2/T_R T_L \sim \omega^2 r_1 r_2$,
in this regime the ``temperature dependent'' corrections are
negligible to the order we are working. Thus, for two large charges
the leading corrections in four and
five dimensions are very different.

\section{Fundamental strings vs.\ black holes}\label{fsvsbh}

{}Fundamental strings can carry two kinds of charges, namely
electric NS-NS charges, associated with the momentum and winding
modes of
the string. Black holes with these same quantum numbers can be
readily
constructed \cite{sen}. Since, in general, string states with given
winding and momentum are highly
degenerate, one would expect a relation with the
Bekenstein-Hawking entropy of the black hole to be viable.
This is not straightforward. For one thing,
a secure starting point for the identification would be
a supersymmetry-protected BPS state, i.e., the extremal black hole.
But for the NS-NS electric black holes
the horizon becomes a zero-area singularity in the extremal limit,
suggesting that stringy corrections to the geometry should be
relevant in its vicinity. Sen  has invoked
these corrections to argue that a
{\it stretched horizon} of string size should be present,
whose area reproduces the degeneracy of
string BPS states \cite{sen2}.

The correspondence principle proposed in \cite{hp}
provides a concrete way to relate strings and black holes (in any
dimension $D\geq 4$) arbitrarily away from extremality,
and correctly obtain, up to factors of
order one, the Bekenstein-Hawking
entropy by counting string states. As explained in \cite{hp},
the comparison between the black hole picture at strong coupling on
the one hand,
and the string to which it evolves
at weak coupling on the other hand,
should be naturally made at the value of the string coupling
where the former description
yields way to the latter. This should happen
when the curvature of the black hole geometry (in the string
conformal frame) reaches the string size. At this point, the
mass and degeneracy of states of the black hole can be matched,
up to factors of order one, to those of the string.

The string to which the black hole evolves is a highly excited one,
in a thermal state. Its decay by annihilation of left and right moving
oscillations reproduces the thermal character of Hawking radiation.
However, the entropy, or equivalently, the Hawking temperature, only
conveys information about the total excitation level of the string. In
particular, it is not possible to tell from the entropy alone any
differences between the excitations of left and
right moving oscillators. Remarkably, as first discussed in \cite{ms},
it turns out that the radiation emitted
from the black hole actually encodes such information in the
greybody filtering of the Planckian spectrum. It is this sort of
analysis what we want to perform here.

Let us, first, briefly review how the correspondence between a
fundamental string and the
black hole goes in the five dimensional case.
Consider a string moving on a circle of radius $R$, carrying $n_p$
and $n_w$ units of momentum and winding, respectively.
If $N_{R,L}$ are the right and
left oscillation level numbers, then the mass levels of the free
string are given by
\begin{equation}\label{strlev}
M^2 = \left( {n_p\over R} + {n_w R\over \alpha'}\right)^2 + 
{4\over \alpha'}N_R
=\left( {n_p\over R} - {n_w R\over \alpha'}\right)^2 + {4\over
\alpha'}N_L.
\end{equation}
A six-dimensional black string can be constructed with these
quantum numbers \cite{hs0}. Its
metric, in the string frame, is
\begin{equation}\label{sixstring}
ds^2_{(6)} =  -{h\over f_w f_p}dt^2
+ {f_p\over f_w}
\left( dz - {r_0^2 \sinh 2\sigma_p\over 2 r^2 f_p}dt\right)^2
+ {dr^2\over h} +r^2 d\Omega_3^2,
\end{equation}
where $f_p,f_w,h$ are functions like those in (\ref{pots}).
The momentum and winding NS-NS charges are identified as
\begin{equation}
n_p = {\pi R\over 8 G_5}r_0^2\sinh 2\sigma_p\;,\qquad
n_w = {\pi \alpha'\over 8 G_5 R}r_0^2\sinh 2\sigma_w.
\end{equation}
The momentum actually results from boosting along $(t,z)$ a string
which initially has $n_p=0$.

We have expressed these charges in terms of the five dimensional
Newton's constant, which can be obtained as
$G_5 = \pi g^2(\alpha')^4/(4 RV), $with $g$ the 10-dimensional
string coupling, and $(2\pi)^4 V$ a compactified
four-volume. The reason for this choice is that upon reduction along
the wrapping direction of the string one obtains
the five dimensional black hole of
section \ref{s5bh} with two non-zero charges.
In our identification with the black hole parameters, we take
$r_w\equiv r_1$
to be the radius associated with winding charge, $r_p\equiv r_2
(\leq r_1)$ associated with the momentum charge
(the case $r_p\geq r_w$ is
$T$-dual to this one), and $r_3=0$.

If we keep the charges fixed, then as we decrease the coupling $g$
the horizon radius $r_0$ becomes smaller. Following \cite{hp},
the string-frame
curvature reaches the string scale when $r_0\sim\sqrt{\alpha'}$.
At this point
the mass of the string in
(\ref{strlev}) can be set equal to that of the black hole
(\ref{5dmass}). This allows one to determine the oscillation levels
$N_R,N_L$.
These could be either
of similar magnitude,
or one much larger than the other.
But remarkably, as noted in \cite{hp},
the sum $\sqrt{N_L}+\sqrt{N_R}$, which corresponds to
the string
degeneracy of states, is, up to a factor
of order one,
independent of the relative size of the summands,
\begin{equation}\label{entstbh}
S_{st}\sim\sqrt{N_L}+\sqrt{N_R}\sim
{(\alpha')^{3/2}\over G_5}\cosh\sigma_w
\cosh\sigma_p \sim S_{bh},
\end{equation}
and therefore
the string entropy correctly reproduces,
within the accuracy of the correspondence principle and for arbitrary
momentum and winding numbers, the
Bekenstein-Hawking entropy of the black hole (\ref{5dmass}) at the
matching point.

Our task now is to test the absorption rates of
scalar particles. The comparison has to be confined to near
extremal situations, where we can use the results of the previous
sections for the classical absorption by the black hole. For 
future reference, the entropy and mass of such black holes 
at the matching point is
\begin{equation}\label{bhcond}
S_{bh} \sim {\alpha'\over G_5} r_w \cosh\sigma_p,\quad
M\sim {r_w^2\over G_5}.
\end{equation}
Before going into the details of the comparison, we must note a number
of peculiarities that arise in the
correspondence between
near extremal black holes and fundamental strings.
The way the extremal limit is reached in the
black hole side requires
sending $r_0^2/G_5\rightarrow 0$ and $\sigma_w\rightarrow\infty$
while keeping $r_0^2\sinh 2\sigma_w/G_5$ (i.e., $n_w$)
fixed. Additionally, one can
also send $\sigma_p \rightarrow\infty$,
keeping $\sigma_w-\sigma_p$ fixed,
thus obtaining an extremal black hole with two charges ($n_w$ and
$n_p$) \footnote{The extremal black hole with a single type
of charge corresponds to a non-degenerate string state, $N_R=N_L=0$,
with zero entropy.
If there are two charges, then $N_R=0$, $N_L\neq 0$.}.
This means that if we want to keep  $r_0 \sim \sqrt{\alpha'}$,
then the extremal limit
corresponds to taking the string coupling $g\rightarrow \infty$.
Nevertheless,
we can still consider near extremal regimes at weak coupling.
The reason is that the mass of the extremal black hole
is $M_{ext} \sim r_w^2/G_5$, whereas the energy above extremality
is $\Delta E\sim r_0^2/G_5$ \footnote{Or, possibly,
$\Delta E\sim r_0^2\cosh 2\sigma_p/G_5$. This does not affect
the argument.}. Then
\begin{equation}\label{closetoext}
{\Delta E\over M_{ext}}\sim {r_0^2\over r_w^2}
\ll 1
\end{equation}
is small independently of the coupling, and it is in this sense
that we talk about a near extremal black hole. Therefore, there is
no problem in a black hole close to extremality
evolving into a weakly coupled string. We should keep in mind, though,
that at weak coupling $\Delta E$ is not infinitesimally small (in
string units) and thus the configuration is a finite distance 
away from the extremal one.

Another important effect in near extremal configurations is that the
gravitational dressing can be
rather large even after the transition to the weakly coupled
string  \cite{hp}.
This comes about by the fact that near the horizon
the factors $f_w$, $f_p$ in the metric
are big ($\sim \cosh^2\sigma_w,\;\cosh^2\sigma_p$), and
thereby induce redshifts in quantities like the compact
radius $R$ or the free energy above the rest mass
with respect to their
asymptotic values.
Nevertheless, Horowitz and Polchinski have argued that the
calculation
of the string entropy (\ref{entstbh}) is not affected by the use of
the asymptotic
values instead of those read from the corrected local metric.

Such redshifts can be
read from (\ref{sixstring}). For example, the local temperature
at the string is related to the temperature measured at asymptotic
infinity as
\begin{equation}
T^{(loc)} = \cosh\sigma_w\cosh\sigma_p T^{(as)}.
\end{equation}
The frequency of quanta emitted by the string undergoes a similar
redshift when they reach the asymptotically flat region. Therefore,
the quotient
\begin{equation}
\left({\omega\over T}\right)^{(loc)}=
\left({\omega\over T}\right)^{(as)},
\end{equation}
which appears in greybody factors, remains unchanged
\footnote{At this point we admit to have found some difficulty
on how to account unambiguosly
for the detailed effect of the redshift when matching black hole
and string parameters. For example, it is not clear to us
how the left and right moving momenta $p_{R,L} = n_p/R\pm
n_w R/\alpha'$ should be redshifted. For the purposes of this 
paper, we will find a simple way to
formulate our arguments that seems to be
free of any such ambiguities.}.

Turn now to analyze the absorption of a neutral scalar of frequency
$\omega$ by a
fundamental string. The
increment in oscillator level due to absorption of neutral quanta
is the same for right and
left movers,
$\delta N_R = \delta N_L$.
Given that the energy increase is $\omega$,
the mass shell condition (\ref{strlev}) yields
\begin{equation}\label{levinc}
\delta N_{R,L} \sim \alpha' M\omega.
\end{equation}
The absorption rate is to be averaged over a statistical ensemble
of initial states peaked at a given mass. The emission rate, from
which the absorption rate can be obtained, has been computed in
\cite{stanford} using string perturbation theory. The result is
\begin{equation}\label{stringabs}
\sigma_{abs}\sim {G_5 (\delta N_L)^2\over \alpha' M\omega}
{e^{\beta_L^*\delta N_L+\beta_R^*\delta N_R}-1 \over
(e^{\beta_L^*\delta N_L}-1)(e^{\beta_R^*\delta N_R}-1)},
\end{equation}
with
\begin{equation}\label{betarl}
\beta_{R,L}^* \equiv {\partial S_{st}\over \partial N_{R,L}}.
\end{equation}
Using (\ref{entstbh}), (\ref{levinc}) and (\ref{betarl}) 
the factors in the
exponentials are
\begin{equation}\label{omegatstring}
\beta_{R,L}^*\delta N_{R,L} \sim {\alpha' M \omega
\over\sqrt{N_{R,L}}}\sim {\omega\over T_{R,L}^{(st)} },
\end{equation}
where the last relation can be taken as a definition of
the left and right moving oscillator effective temperatures.
\footnote{Since eventually we are only interested in comparing the
functional form of the greybody factors, we do not need to interpret
$T_{R,L}^{(st)}$ as actual temperatures. In fact, we could even do
without them. But recall that for a weakly coupled closed string
the left and right moving oscillations behave much independently of
each other, being only related by the mass shell condition.
The total degeneracy of the string
is the sum of the degeneracies $S_{R,L} \sim
\sqrt{N_{R,L}}$ of
each separate ensemble, so we could define effective left and right
temperatures by $1/T_{R,L}^{(st)}\sim (\partial S_{R,L}/ \partial
N_{R,L})(\delta N_{R,L}/\delta M)$.
These coincide with (\ref{omegatstring}).}
It is important to notice that the frequency here is measured
at the {\it local} position of the string. Similarly,
$T_{R,L}^{(st)}$ are local quantities as well.

An assumption needed to derive the result (\ref{stringabs}) is that
the Compton
wavelength of the scalar be much bigger than the
string scale,
\begin{equation}
\omega\sqrt{\alpha'} \ll 1.
\end{equation}
At the matching point, $r_0\sim\sqrt{\alpha'}$,
this condition is, in fact, less restrictive than the one
imposed on
the semiclassical calculation.

Consider first the leading term in the very low frequency limit. By
expanding (\ref{stringabs}), and using the expressions for
$\delta N_{R,L}$ (\ref{levinc}), (\ref{omegatstring}), one finds
\begin{eqnarray}
\sigma_{abs} &\sim& G_5 { \beta_R^* +\beta_L^*
\over \beta_R^* \beta_L^*} \sim G_5 \left(\sqrt{N_R}
+\sqrt{N_L}\right)\nonumber\\
&\sim& A_{bh},
\end{eqnarray}
thus correctly reproducing the semiclassical leading order result.
The proportionality, with a factor of order one, between
$\sigma_{abs}$ and the black hole area was found in \cite{stanford}.
The fact that the agreement is not precise is something that the
correspondence principle allows for.

As mentioned above, in the quantities tested so far only the
particular combination of oscillator level numbers that yields
the entropy, or area, enters, so that we have not been able
to discern the
individual values of the left and right moving oscillator
levels. The greybody factors, which depend solely on the quantities
$\omega/T_{R,L}$ can convey such information. In this respect,
the formal similarity
between the perturbative string and classical
black hole results, (\ref{sigmaphys}) and (\ref{stringabs})
is most remarkable.
Unfortunately, we can
easily see that the left and right moving temperatures read from the
black hole absorption spectrum can not agree with those obtained
from the string
spectrum. To this effect, first we must redshift the asymptotic
temperatures (\ref{lrtemps}) (with $r_1=r_w$, $\sigma_2=\sigma_p$,
and $\sigma_3=0$) to the
location of the string $r\sim r_0\sim \sqrt{\alpha'}$. This yields
\begin{equation}
T^{(loc)}_{R,L}\sim {1\over \sqrt{\alpha'}},
\end{equation}
i.e., both temperatures are of the order of the Hagedorn temperature.
Suppose now that they were equal, up to factors of order one,
to the local string temperatures obtained
from (\ref{omegatstring}). This would translate into
$\sqrt{N_R}\sim \sqrt{N_L} \sim \sqrt{\alpha'} M$,
for any values of
the charges (requiring only large $\sigma_w$).
But if we take (\ref{bhcond}) into account, we find that {\it this
condition is
incompatible with the entropy being given by $S \sim \sqrt{N_R} +
\sqrt{N_L}$}.

The conclusion follows that temperatures can not
match and therefore string and black hole greybody factors
disagree in their functional dependence.
The discrepancy is most patent when
$r_w\gg r_p\gg 1$,
but is true throughout virtually all of the parameter range
we can probe\footnote{It is perhaps
worthwhile noting that the disagreement would also be present
if we had not redshifted the asymptotic temperatures to their local
values.}.
Additionally, when there is essentially one large charge,
and $\omega/T$
becomes small, the logarithmic corrections in (\ref{sigmaphys2})
become another source of trouble.

The same sort of discrepancy appears when comparing the left and
right moving
temperatures for the four dimensional black hole, though in this
case the parameter range is more restricted. Finally, the 
wave equation for black holes in dimensions higher than five
can not be solved, close to the horizon, in terms of
hypergeometric functions, and therefore
we do not know how to obtain greybody factors such as those in
(\ref{sigmaphys}). However, in light of the results we have found it
would not be surprising if the disagreement persisted in $D\geq 6$.

\section{Discussion}
\label{concl}

Several comments are in order regarding the discordant result found
in the previous section. First of all, one should note
that it does not disprove the correspondence principle. At the low
frequencies we are working, the greybody corrections to the
absorption rates are either of order one (when there are two large
charges and (\ref{sigmaphys}) is valid) or subleading
(with just one large charge, (\ref{sigmaphys2})). The modulation in
frequency in the former case is not big.  Therefore, as far as the
absorption rates are concerned, there is no large change throughout
the transition from the black hole to the weakly coupled string.

Nevertheless, our result places a limit on the applicability of
the simplest model for the correspondence, which thus becomes
comparatively less powerful
than the ``effective string'' model for D-brane black holes.
It appears that the low frequency corrections to the
absorption rates
undergo seemingly significant changes
in the transition from the black hole to the string.
One would expect a more detailed examination
of the correspondence to reveal the reason.
Apparently, the lowest order perturbative string result (amplitudes
on the sphere) can
not account for this. One-loop corrections to the
closed string vertex should add terms
$\propto g^2$.
On the other hand, in the regime where only one of the
charges is large, the
correction term resulting from expanding the greybody factors is
\begin{equation}
{\omega^2\over T_R T_L} \propto g^2\omega^2,
\end{equation}
but recall that this is shadowed by the logarithmic correction
in (\ref{sigmaphys2}) (which is $\propto g^2$, too).
Things are further complicated by the fact that
in four dimensions the leading term at small frequencies goes like
$g^2\omega$.
It is very unclear whether string vertex
corrections can simultaneously account for all these facts.

Apart from the perturbative corrections
(higher powers of $g$) to the emission
spectrum of the string, another source for possible corrections of the
correspondence comes from string-size ($\alpha'$) effects
on the black hole geometry. However, these do not seem to be
important to account for the
area (entropy) of the black hole horizon (this is part of the content
of the correspondence principle), so it is uncertain whether
they might significantly alter the role of
the horizon as a boundary for scattering wavefunctions. Also, such
corrections are presumably very hard to compute.

Actually, there are reasons to suspect that the NS-black hole/string
correspondence might not be so simple for black holes close to
extremality. From what we have seen, if we keep the horizon at
a string scale size,
the BPS limit is only reached
for $g\rightarrow \infty$. Thus, a near extremal black hole
that evolves
into a {\it weakly} coupled string might {\it not} be close to
the BPS state $N_R=0$.
This is in contrast with the situation for the black hole with
three nonvanishing charges \cite{callan,hs,ms}, where a state can
remain infinitesimally close to extremality throughout the passage
from strong to weak coupling. For the case we study here, with only
two charges, the weakly coupled state is instead always a finite 
distance away from the extremal state.

This is very presumably related to the lack of agreement of radiation
profiles. Consider a string
that is close to the BPS state; we would expect this condition to mean
$N_L\gg N_R$. In turn, this would mean $T_L^{(st)} \gg T_R^{(st)}$,
which can hardly be harmonized with the
classical black hole result,
which requires $T_L = T_R$ for all values
of the parameters (near extremality).
Indeed, it is not likely that $T_L \gg T_R$ can hold at the
transition point. The reason is that the Hawking
temperature is related to the left and right temperatures by
\begin{equation}
{2\over T_H} = {1\over T_L} + {1\over T_R}.
\end{equation}
A similar relation also holds for the temperatures defined from the
string spectrum.
It implies that, if one of the temperatures is much
higher than the other, say, $T_L \gg T_R$, then $T_R \sim T_H\ll T_L$.
But, as we have seen, the gravitational dressing turns the asymptotic
temperature of the radiation $T_H$ into a local string temperature
$1/\sqrt{\alpha'}$ at the matching point, and therefore
we would have $T_L^{(loc)}$ much larger than the string scale,
which does not seem reasonable.
A more conceivable scenario would be that, as
the black hole shrinks to the string scale, the number of left moving
oscillators $N_L$ can never reach a value much bigger than $N_R$,
in the sense
that we do not get $N_L/N_R \gg 1$, though the
difference $N_L-N_R =n_p n_w$, which is kept fixed,
can still be quite large. Such a string would not be close
to the BPS state. It would have $T^{(st)}_R\sim T^{(st)}_L\sim 
1/\sqrt{\alpha'}$, and
could possibly agree with the left and right black hole temperatures.

There does not appear to be any simple enough way to 
implement this picture. A complete analysis should involve a proper 
treatment of the locally corrected quantities. However, this 
is unlikely
to be enough for solving the puzzle, at least within the simplest
model for black hole/string correspondence (i.e., that without
higher order corrections to the string vertex or the
black hole geometry).
The reason is that our result in the previous section
seems to be largely independent of how the gravitational
dressing acts.
The redshift effect does not seem to suffice to obtain
$T_L^{(st)} \sim T_R^{(st)}$, and {\it simultaneously} preserve the
agreement between string and black hole entropies. 
The conclusion seems to be that, at least in this region of 
black hole parameter space, the detailed correspondence to 
fundamental strings is not simple.

As an aside, in view of the results above, we find it remarkable,
though somewhat puzzling, that if these same
fundamental strings are placed in the background of a
magnetic NS fivebrane, then
{\it precise} agreement can be found both for the entropies and the
emission rates \cite{hal}. Apparently, the only effect of the
fivebrane on the string
is to rescale the oscillator number and string tension, and restrict
the motion of the string to the worldvolume of the fivebrane.
Otherwise, the dynamics of the string is unaltered. The resulting
black hole, on the other hand, possesses now a regular
horizon in the extremal limit. The correspondence
principle, however, cannot be applied to these configurations, since
their horizon does not decrease below the string size at weak
coupling.

It is perhaps disappointing to find that the simplest approximation
to the black hole/string correspondence does not seem to work as well
as it does for other kinds of black holes, namely, those with
a regular extremal limit.
There are number of other sorts of scalars
(charged, fixed, intermediate \cite{klebetal}) 
that can be used to probe further the
correspondence between black holes and fundamental strings, and may
shed further light on the problem. This is
currently under investigation.

\section*{Acknowledgements}
We are indebted to Gary Horowitz and John Pierre for useful 
conversations on these issues. We also thank 
Arvind Rajaraman for pointing out an error in a
previous version of this paper and for making his
notes available \cite{rajaraman}. 
This work has been partially 
supported by a postdoctoral FPI fellowship (MEC-Spain) and by grant 
UPV 063.310-EB225/95.

\end{document}